\documentclass[12pt]{article}
\usepackage{pslatex}
\usepackage{txfonts}
\usepackage{cite}
\usepackage{color,colordvi}
\usepackage{graphicx,scalefnt}

\def\ca{{C^{}_A}}
\def\cf{{C^{}_F}}

\def\mt{{m_t}}
\def\mts{{m_t^2}}

\def\muf{{\mu^{}_f}}
\def\mufs{{\mu^{\,2}_f}}
\def\mur{{\mu^{}_r}}
\def\murs{{\mu^{\,2}_r}}

\def\gE{{\gamma_{E}}}
\def\z#1{{\zeta_{#1}}}

\def\lnN{\ln N}
\def\lnNs{\ln^{\,2} N}

\def\b#1{\beta_#1}
\def\Aq#1{A^{(#1)}_q}
\def\DQQ#1{D^{(#1)}_{Q{\bar Q}}}

\newcommand{\Lqf}{\ln\left(\frac{M^2}{\mu_f^{\,2}}\right)}

\newcommand{\Lqr}{\ln\left(\frac{M^2}{\mu_r^{\,2}}\right)}

\newcommand{\Lfr}{\ln\left(\frac{\mu_f^{\,2}}{\mu_r^{\,2}}\right)}

\def\MSbar{\ensuremath{\overline{\mbox{MS}}}}

\newcommand{\gsim}{\;\rlap{\lower 3.5 pt \hbox{$\mathchar \sim$}} \raise 1pt
 \hbox {$>$}\;}
\newcommand{\lsim}{\;\rlap{\lower 3.5 pt \hbox{$\mathchar \sim$}} \raise 1pt
 \hbox {$<$}\;}

\def\JPC{\mbox{J}^{\mbox{\scriptsize PC}}}
\def\Im{\mbox{Im}}
\def\mt{m_t}
\def\cA{{\cal A}}

\def\cAc{{\cal A}_{c}}
\def\cAnc{{\cal A}_{nc}}

\def\cCh{{{\cal C}_h}}

\begin{document}


\title{\vskip-3cm{\baselineskip14pt
    \begin{flushleft}
      \normalsize DESY 08-169\\
      \normalsize HU-EP-08/57\\
      \normalsize SFB/CPP-08-94\\
      \normalsize TTP08-52
  \end{flushleft}}
  \vskip1.5cm
  Top-quark pair production near threshold at LHC
}
\author{\small Y. Kiyo$^{(a)}$,
  J.H. K\"uhn$^{(a)}$,
  S. Moch$^{(b)}$,
  M. Steinhauser$^{(a)}$,
  P. Uwer$^{(c)}$\\[1em]
  {\small\it (a) Institut f{\"u}r Theoretische Teilchenphysik,
    Universit{\"a}t Karlsruhe (TH)}\\
  {\small\it Karlsruhe Institute of Technology (KIT)}\\
  {\small\it 76128 Karlsruhe, Germany}
  \\
  {\small\it (b) Deutsches Elektronen-Synchrotron DESY}\\
  {\small\it D-15738 Zeuthen, Germany}
  \\
  {\small\it (c) Institut f\"ur Physik, Humboldt-Universit\"at zu Berlin,}\\
  {\small\it D-10099 Berlin, Germany}
}

\date{}

\maketitle

\thispagestyle{empty}

\begin{abstract}
The next-to-leading order analysis for the
cross section for hadroproduction of top quark pairs close to
threshold is presented. Within the
framework of non-relativistic QCD a significant enhancement
compared to fixed order perturbation theory is observed
which originates from the characteristic remnant of the $1S$ peak
below production threshold of top quark pairs.
The analysis includes all color singlet and color octet configurations 
of top quark pairs in $S$-wave states and, for the dominant configurations, 
it employs all-order soft gluon resummation for the hard parton cross section.
Numerical results for the Large Hadron Collider at $\sqrt{s} = 14~$TeV 
and $\sqrt{s} = 10~$TeV and also for the Tevatron are presented.
The possibility of a top quark mass measurement from the invariant 
mass distribution of top quark pairs is discussed.
\medskip

\end{abstract}


\thispagestyle{empty}

\newpage



\section{Introduction}

At the CERN Large Hadron Collider (LHC) the major part of top
quarks are produced in pairs. Due to the experience gained at the
Fermilab Tevatron \cite{ICHEP:2008} and the huge amount of top quarks
to be produced at LHC the reconstruction of
top quarks with  good accuracy will be
possible~\cite{atlastdr:1999,cmstdr:2006}.
A significant fraction of top quark pairs will be
produced close to threshold. Thus a dedicated analysis of the production
cross section in this region is required which is best performed
within the framework of non-relativistic QCD
(NRQCD)~\cite{Bodwin:1994jh,Bodwin:1994jherr}. 

The production of top anti-top quark pairs close to the
kinematical threshold has received much attention in the 
context of precision measurement of top quark properties at 
a future International Linear Collider (ILC).
Theoretical calculations and dedicated experimental analyses
have demonstrated that a precise extraction of the top quark mass, 
its width and the strong coupling constant is
possible{~\cite{Martinez:2002st,Fujii:1993mk}} at the ILC. 
The complete next-to-next-to-leading order (NNLO) predictions are
available since many years~\cite{Hoang:2000yr}. (For earlier work
see e.g \cite{Fadin:1987wz,Strassler:1990nw,Jezabek:1992np,Sumino:1992ai}.)
Partial
next-to-next-to-leading logarithmic
(NNLL)~\cite{Hoang:2003xg,Pineda:2006ri} and
next-to-next-to-next-to-leading order
(NNNLO)~\cite{Beneke:2007gj,Beneke:2007pj,Beneke:2008cr}
predictions were evaluated more recently.

In contrast to the linear collider, where the physical observable is
the total cross section as a function of energy, at the hadron collider
one considers the invariant mass distribution of the top quark pairs.
Since the expected uncertainty is significant larger than the
one anticipated at a linear collider a next-to-leading order (NLO)
analysis is probably sufficient.
The calculation of the cross section within the NRQCD framework 
contains as building blocks the
hard production cross section for a top quark pair at threshold and the
non-relativistic Green's function governing the dynamics of the would-be
boundstate. Both ingredients are available in the literature since
many years. In particular, the
hard cross section for threshold $t\bar{t}$ production can be found
in Refs.~\cite{Kuhn:1992qw,Petrelli:1997ge}.
In Ref.~\cite{Kuhn:1992qw} the NLO formulae were derived for quark or
gluon initial states and a quarkonium in a $\JPC=0^{-+}$ color singlet
state, plus possibly a parton. The general case, with the heavy quark system 
$(Q\bar Q)$ in $S$-wave singlet/triplet spin state, and color singlet/octet
configuration 
is given in Ref.~\cite{Petrelli:1997ge}, together with the corresponding
results for $P$-waves.
The results of Refs.~\cite{Kuhn:1992qw,Petrelli:1997ge} were presented
for stable boundstates. For unstable wide resonances it is convenient
to describe the bound state dynamics through a Green's function.

Recently a calculation of top quark threshold hadroproduction near threshold
has appeared~\cite{Hagiwara:2008df}. (For an early discussion along similar
lines see~\cite{Fadin:1990wx}.)
The basic idea of our approach is
similar to the one of Ref.~\cite{Hagiwara:2008df}. We aim a detailed study 
of the top quark production based on NLO cross section formulae in the 
NRQCD framework. In our set-up all NLO sub-processes have been included, 
i.e., also those which appear for the first time in ${\cal O}(\alpha_s^3)$. 
{Furthermore, the matching between QCD and NRQCD as performed in
Ref.~\cite{Hagiwara:2008df} and the present paper is slightly
different. Whereas in~\cite{Hagiwara:2008df} the matching has been performed
for the limit where the partonic center-of-mass energy $\hat{s}$ approaches
twice the top quark mass we include the complete dependence on $\hat{s}$ as
given in Refs.~\cite{Kuhn:1992qw, Petrelli:1997ge}. Thus, formally, the result
of Ref.~\cite{Hagiwara:2008df} is only valid for top-quark production where
the velocity of both quarks is small. On the other hand, in our approach 
the relative velocity has to be small whereas the top-anti-top quark system
can still move with high velocity.
}
Finally, we perform a soft gluon resummation which enhances
the cross section by a few per cent.

Our paper is organized as follows: In the next
Section details of the formalism  used for the calculation
of the NLO cross section are provided.
The effects of initial-state radiation and
the hard contribution are discussed in Section~\ref{sec::hard} and the
soft gluon resummation is performed in Section~\ref{sec::resum}.
The properties of the Green's function are summarized in
Section~\ref{sec::bound}. In Section~\ref{sec::results} the
building blocks are combined and numerical results for the invariant mass
distribution are presented. Theory uncertainties
due to scale variation and unknown higher order corrections are
estimated. Summary and conclusions are
presented in Section~\ref{sec::summary}.


\section{The production cross section}

Let us denote the (quasi) boundstate of a top and anti-top
quark  with spin $S$ and angular momentum $L$
by $T\equiv{^{2S+1}L}_J^{[1,8]}$ where the superscripts
$[1]$ and $[8]$ denote the singlet and octet color states.
The production rate is obtained from the production cross 
section of a top quark pair
with invariant mass $M^2\equiv (p_t+p_{\bar{t}})^2$ 
and its evolution to a quasi boundstate
described by the non-relativistic QCD.
The former is a hard QCD process at a distance $\sim 1/m_t$ and thus
computable within the conventional perturbative expansion in $\alpha_s$.

The long-distance effects responsible for the formation of a narrow
boundstate are described by the squared wave function at the origin
$|\Psi(0)|^2$ or, in the language of NRQCD, by the matrix elements
\begin{equation}
  \langle\,(\chi^\dagger\,\Gamma\,\psi)
  \,\cdot (\psi^\dagger\Gamma\,\chi)\,\rangle =
  {\cal N}_s\, {\cal N}_c\, |\Psi(0)|^{\,2}.
\end{equation}
Here ${\cal N}_s = 2 S +1$ and ${\cal N}_c = 1, 8$ denote the number of spin and
color degrees of freedom, respectively.
We are interested in the differential distribution ${\rm d}\sigma/{\rm d}M$
which, for narrow resonances with mass $M_n$, is proportional to 
$\delta(M-M_n)$. 
For wide resonances, the case under consideration, it is convenient to
convert the factor describing the sum over individual resonances into
the non-relativistic Green's function\footnote{In the case of color octet
  states we cannot take Eq.~(\ref{eq:WFtoGF}) literally but derive 
  a corresponding formula within the framework of NRQCD.
}
\begin{eqnarray}
  \sum_n |\Psi_n(0)|^2\,\pi\,\delta(M-M_n) 
  &\to&
  \sum_n\, \Im {\Psi_n(0)\Psi_n^\ast(0)\over M_n - (M+i\Gamma_t)
  }
  = \Im \,G(M+i \Gamma_t),
  \label{eq:WFtoGF}
\end{eqnarray}
with $G(M+i\Gamma_t)\equiv G^{[1,8]} (\vec{r}=0; M+i \Gamma_t)$ being the
Green's function at zero distance for the nonrelativistic Schr\"odinger equation 
discussed below.
Since the typical momentum scale governing the nonrelativistic top quark system 
$\mt \varv$ (with $m_t \varv^2 \equiv M+i\Gamma_t-2m_t$, and $\varv$ being the 
velocity of top and anti-top quarks) is in the perturbative regime,
and the large top quark width $\Gamma_t$ introduces an additional cutoff
scale $\sqrt{m_t\Gamma_t}$, the Green's
function can be evaluated perturbatively. 
As stated above the present paper is concerned with the production of
top quark pairs near threshold, thus restricted to states with $L=0$,
i.e. $T=\,{^{2S+1}S}_J^{[1,8]}$. The contributions to the invariant
mass distribution with higher angular momentum are at least suppressed 
by $\varv^{2}$, and thus of higher order (beyond NLO).

In order to obtain experimentally measurable quantities at a hadron
collider the partonic differential cross section
${{\rm d}\hat\sigma_{ij\rightarrow T}/{\rm d}M}$ 
is convoluted with the luminosity function
\begin{eqnarray}
  \bigg[\frac{d{\cal L}_{ij}}{d\tau}\bigg](\tau,\mufs)
  &=&
  \int_0^1 {\rm d}x_1 \int_0^1 {\rm d}x_2\, 
  f_{i/P_1}(x_1,\mufs) f_{j/P_2}(x_2, \mufs)\,
  \delta(\tau - x_1 x_2)
  \,,
  \label{eq:lumi}
\end{eqnarray}
where $i,j$ refer to partons inside the hadrons $P_1$ and $P_2$ with the 
distribution functions $f_{i/P_1}$ and $f_{j/P_2}$.
The dependence on the factorization
scale $\mu_f$ cancels in combination with the one contained
in ${\rm d}\hat{\sigma}_{ij\rightarrow T}/{\rm d}M$.
The differential cross section can thus be written as
\begin{equation}
 M {{\rm d}\sigma_{P_1 P_2\to T}\over {\rm d}M}(S, M^2) 
 =\sum_{i,j} \int_\rho^1 {\rm d}\tau\,
  \bigg[\frac{{\rm d}{\cal L}_{ij}}{{\rm d}\tau}\bigg](\tau,\mufs)
  ~M\frac{{\rm  d}\hat\sigma_{ij\to T}}{{\rm d}M}(\hat s,M^2,\mufs)\,.
 \label{eq:TheConvolution}
\end{equation}
As usual $\hat{s}$ and $S$ denote the partonic and 
the hadronic center-of-mass energy squared, respectively, 
and $\tau=\hat s/S$.
The lower limit of the $\tau$ integration is given by $\rho=M^2/S$.
The partonic differential cross section ${\rm d}\hat \sigma_{ij\to T}/{\rm d}M$
consists of a factor $F$ that is evaluated in perturbative QCD,
and can be deduced from Refs.~\cite{Kuhn:1992qw,Petrelli:1997ge},
and a second factor, the imaginary part of the Green's function $G^{[1,8]}$
\begin{eqnarray}
  M\frac{{\rm d}\hat\sigma_{ij\to T}}{{\rm d}M}(\hat s,M^2,\mufs)
 &=&
 F_{ij\to T}(\hat{s},M^2,\mufs)\,
\, \frac{1}{m_t^2}\,{\rm Im} \,G^{[1,8]}(M+i\Gamma_t)\, ,
\label{eq:PartonicXS}
\end{eqnarray}
where the superscript of the Green's function refers to the color state of $T$. 
Eqs.~(\ref{eq:TheConvolution}) and (\ref{eq:PartonicXS}) constitute our
master formulae, which contain
several scales and various physics contributions
of different origin in factorized form. In particular,
the soft dynamics of the parton distribution and real
radiation is contained in the convolution of $F_{ij\to T}$
with the parton luminosity, the boundstate effects are described by $G$.
Note that at NLO the Green's function $G^{[1,8]}(M+i\Gamma_t)$ and the convolution
of $F$ with the parton luminosity $({\cal L}\otimes F)$
are individually independent of the renormalization 
scale $\mur$.
Thus we can discuss the two parts separately in the
following two Sections. Furthermore, it is simpler to assess the uncertainties
for the individual contributions.

Let us at this point make a comment concerning the
validity of {Eq.~(\ref{eq:PartonicXS}),} 
which makes use of the NRQCD
expansion assuming $\varv\ll 1$, thus being limited to the
threshold region.
For larger invariant masses  conventional perturbation theory is
applicable (see
Refs.~\cite{Nason:1988xz,Beenakker:1989bq,Bernreuther:2004jv}
and Refs.~\cite{Moch:2008qy,Moch:2008ai,Cacciari:2008zb,Kidonakis:2008mu} 
for recent compilations
{of the total cross section and~\cite{Frederix:2007gi} for a proposal
to measure the top-quark mass from the shape of
${\rm d}\sigma/{\rm d} M$}).
In the transition region the predictions from both methods are
expected to coincide, as will be discussed below
(c.f. Fig.~\ref{fig:mtt_threshold}).


\section{Hard cross section}
\label{sec::hard}

In this Section the ingredients for the NLO corrections to the hard
cross section will be collected, which are taken from
Refs.~\cite{Kuhn:1992qw,Petrelli:1997ge}.
We parameterize the function $F_{ij\to T}$, representing the hard cross section 
for $ij\rightarrow T X$ 
($X$ stands for additional partons in the inclusive cross sections),
in the following form:
\begin{eqnarray}
  F_{ij \rightarrow T}(\hat s, M^2,\mufs)
  &=&
  {\cal N}_{ij\to T} \,
  \frac{\pi^2\,\alpha_s^2(\mur)}{3\hat s}\,
  \left( 1 + \frac{\alpha_s(\mur)}{\pi}\, {\cCh} \right)
\nonumber \\
&& \hspace{-0.4cm}
  \times
  \left[\, \delta_{ij \rightarrow T} \, \delta (1-z)
    +
    \frac{\alpha_s(\mur)}{\pi}\,
    \bigg( \cAc(z) + \cAnc(z) \bigg)\,
  \right]\,.
\label{eq:F}
\end{eqnarray}
Here $\delta_{gg\rightarrow {^1S_0^{[1,8]}}}
=\delta_{q\bar{q}\rightarrow {^3 S_1^{[8]}}}=1$ and zero for all other
$2\to1$ processes, and $z=M^2/\hat{s}$.
The quantities $\cAc$, $\cAnc$, and $\cCh$
all depend on $i$, $j$, and $T$, the functions $\cA$ in addition on $z$.

The coefficients $\cCh$ originate from the hard corrections to the
production process.
The functions $\cAc$ contain the real corrections with collinear
parton splitting from one of the initial partons $i,j$, and are
governed by the Altarelli-Parisi splitting functions,
$\cAnc$ originates from non-collinear real emission. These
individual contributions are manifest already in Ref.~\cite{Kuhn:1992qw} 
and the appendix of Ref.~\cite{Petrelli:1997ge}, and will be listed in the
following. 
Note, that in Eq.~(\ref{eq:F}) we have split off the factor
$(1+(\alpha_s/\pi)\,\cCh)$, which we 
attribute to hard corrections and thus
treat as a multiplicative factor
to the terms in square brackets. 

\begin{table}
\begin{center}
{
\scalefont{0.98}
\renewcommand{\arraystretch}{1.4}
\begin{tabular}{c|c|c|c|c|c}
\hline
  $gg\rightarrow {^1S_0}^{[1,8]}$
& $gq\rightarrow {^1S_0}^{[1,8]}$
& $q\bar{q}\rightarrow {^1S_0}^{[1,8]}$
& $gg\rightarrow {^3S_1}^{[1,8]}$
& $gq\rightarrow {^3S_1}^{[1,8]}$
& $q\bar{q}\rightarrow {^3S_1}^{[1,8]}$
\\
\hline
\hline
  $ \big[\, 1, \,5/2 \,\big]$
& $ \big[\, 1, \,5/2\,\big]$
& $ \big[\, 3/4, \,6\,\big]$
& $ \big[\, 9/4, \,18\,\big]$
& $ \big[\, 0, \,32/3\,\big]$
& $ \big[\, 0, \,32/3\,\big]$
\\
\hline
\end{tabular}
\renewcommand{\arraystretch}{1.0}
}
\caption{Normalization factors ${\cal N}_{ij\to T}$  
for each process for [singlet, octet] color states.
($N_c=3$ is used.) 
}
\label{tb:XSnorm}
\end{center}
\end{table}

In Tab.~\ref{tb:XSnorm} we collect all processes of the type
$ij \rightarrow TX$ at NLO which contribute in our analysis and
list the corresponding normalization factors ${\cal N}_{ij\to T}$.
Note that the production of a spin triplet color singlet state ${^3}S_1^{[1]}$
via $gq$ or $q\bar{q}$ scattering is zero
up to and including NLO. This is because in these channels the heavy quarks are produced
through gluon splitting $g^\ast \rightarrow t \bar{t}$, which
is only possible if the $t\bar{t}$ is in an octet state.

The coefficients $\cCh$ are non-vanishing only for the
processes which are present also in lowest
order{~\cite{Kuhn:1992qw,Petrelli:1997ge}}: 
\begin{eqnarray}
  {\cal C}_h[gg\rightarrow{^1S_0^{[1]}}]
  &=&
  \frac{\beta_0}{2}\ln\left(\frac{\murs}{M^2}\right)
  + C_F\left(\frac{\pi^2}{4}-5\right)
  +C_A\left(1+\frac{\pi^2}{12}\right)
  \,,\nonumber\\
  {\cal C}_h[gg\rightarrow{^1S_0^{[8]}}]
  &=&
   \frac{\beta_0}{2}\ln\left(\frac{\murs}{M^2}\right)
  +C_F\left(\frac{\pi^2}{4}-5\right)
  +C_A\left(3-\frac{\pi^2}{24}\right)
  \,,\nonumber\\
  {\cal C}_h[q\bar{q}\rightarrow{^3S_1^{[8]}}] 
  &=&
   \frac{\beta_0}{2}\ln\left(\frac{\murs}{M^2}\right)
  +C_F\left(\frac{\pi^2}{3}-8\right)
  +C_A\left(\frac{59}{9}+\frac{2\ln{2}}{3}-\frac{\pi^2}{4}\right)
\nonumber\\&&  
  -\frac{10}{9} {n_f} T_F -\frac{16}{9}T_F
  \,,
\label{eq:cCh}
\end{eqnarray}
where $\beta_0=(11/3)\, C_A  -(4/3)\, n_f T_F$ and 
$C_F=4/3, ~C_A=3,~ T_F=1/2,~ n_f=5$.
The last term in ${\cal C}_h[q\bar{q}\rightarrow{^3S_1^{[8]}}]$,
arising from non-decoupling of the top quark in the gluon propagator,
has been observed and discussed in Ref.~\cite{Hagiwara:2008df}, {see
  also footnote~3 on page~73 of Ref.~\cite{Hagiwara:2008df}}.
For the other processes hard corrections are of higher order, thus $\cCh$
is zero at NLO:
\begin{eqnarray}
  &&
   {\cal C}_h[gq\rightarrow{^1S_0^{[1,8]}}]
  ={\cal C}_h[q\bar{q}\rightarrow{^1S_0^{[1,8]}}]
  ={\cal C}_h[gg\rightarrow{^3S_1^{[1,8]}}]
  ={\cal C}_h[gq\rightarrow{^3S_1^{[8]}}]
  =0.
\end{eqnarray}

The function ${\cal A}_c$ is conveniently expressed using Altarelli-Parisi
splitting functions $P_{ij}(z)$ introduced
below{~\cite{Kuhn:1992qw,Petrelli:1997ge}}
\begin{eqnarray}
\cAc[gg\to {^1S_0^{[1,8]}}] 
&=&
(1-z) P_{gg}(z)\,\left\{
2\,\left[ \frac{\ln(1-z)}{1-z}\right]_+
+ \left[\frac{1}{1-z}\right]_+\, \ln\left(\frac{M^2}{z\,\mu_f^2}\right)
\right\}
-\frac{\beta_0}{2}\, \delta(1-z)\,\ln\left(\frac{\mu_f^2}{M^2}\right)
\,,\nonumber\\
\cAc[gq\to {^1S_0^{[1,8]}}] 
&=&
\frac{1}{2}\,P_{gq}(z)\,
\ln\left(\frac{M^2\,(1-z)^2}{z\,\mu_f^2}\right)
+\frac{C_F}{2}\,z
\,,\nonumber\\
\cAc[q\bar{q}\to {^1S_0^{[1,8]}}] 
&=& 
0
\,,\nonumber \\
\cAc[gg\to {^3S_1^{[1,8]}}] 
&=&
0
\,,\nonumber \\
\cAc[gq\to {^3S_1^{~[8]}}] 
&=&
\frac{1}{2}\,P_{qg}(z)\,
\ln\left(\frac{M^2\,(1-z)^2}{z\,\mu_f^2}\right)
+ T_F\,z\, (1-z)
\,,\nonumber \\
\cAc[q\bar{q}\to {^3S_1^{~[8]}}] 
&=&
(1-z)\, P_{qq}(z)\,\left\{
2\,\left[ \frac{\ln(1-z)}{1-z}\right]_+
+ \left[\frac{1}{1-z}\right]_+\, \ln\left(\frac{M^2}{z\,\mu_f^2}\right)
\right\}
+ C_F\,(1-z)
\nonumber\\&&\mbox{}
-\frac{3\,C_F}{2}\,\delta(1-z)\,\ln\left(\frac{\mu_f^2}{M^2}\right)
\,,
\label{eq:cAc}
\end{eqnarray}
where the conventional plus-distribution\footnote{
The plus-distribution follows the prescription
$ 
\int_{0}^1 dz\, \big[\frac{\ln^n(1-z)}{1-z}\big]_{+} f(z)
\equiv\int_{0}^1 dz\, \frac{\ln^n(1-z)}{1-z} \big[ f(z)-f(1) \big],
$
where $f(z)$ is an arbitrary test function which is regular at $z=1$.
It is related to the $\rho$-prescription used in Ref.~\cite{Petrelli:1997ge} by
$
\big[{\ln^n(1-z)\over 1-z}\big]_+
  = \big[{\ln^n(1-z)\over 1-z}\big]_\rho
  + {\ln^{n+1}(1-\rho)\over n+1}\,\delta(1-z).
$
}
was employed to regularize the singularity at $z=1$. 
The splitting functions $P_{ij}(z)$ are given by
\begin{eqnarray}
P_{gg}(z)
&=&
2\,C_A\bigg[\frac{1}{1-z}+\frac{1}{z}+z(1-z)-2\bigg]\,,
\nonumber \\
P_{gq}(z)
&=&
C_F\bigg[\frac{1+(1-z)^2}{z}\bigg]\,,
\nonumber\\
P_{qg}(z)
&=&
T_F\bigg[z^2+(1-z)^2\bigg]\,,
\nonumber\\
P_{qq}(z)
&=&
2\,C_F\bigg[\frac{1}{1-z}-\frac{1+z}{2}\bigg]\,.
\label{eq:APSF}
\end{eqnarray}
The functions $\cAnc$ are obtained from the non-collinear
contributions. For spin singlet states we have
\begin{eqnarray}
\cAnc[gg\to {^1S_0^{[1]}}] 
&=&
\frac{-C_A}{6 z (1-z)^2 (1+z)^3}
\bigg[
12+11 z^2+24z^3-21z^4-24z^5+9z^6
\nonumber\\ &&\mbox{}
-11z^8+12\left(-1+5z^2+2z^3+z^4+3z^6+2z^7\right)\ln{z}
\bigg]
\,,\nonumber\\
\cAnc[gg\to {^1S_0^{[8]}}] 
&=&
\frac{-C_A}{6z\,(1-z)\,(1+z)^3}
\bigg[
12+23z^2+30z^3-21z^4-24z^5+9z^6-6z^7
\nonumber\\ &&\mbox{}
-23z^8
+\left(-12+60z^2+24z^3+36z^4+60z^6
+24z^7\right)\ln{z}
\bigg]\, 
\bigg(\frac{1}{1-z}\bigg)_+
\,,\nonumber\\
\cAnc[gq\to {^1S_0^{[1,8]}}] 
&=& 
-C_F\,\frac{1}{z}\,\left(1-z\right)\,(1-\ln{z})
\,,\nonumber \\
\cAnc[q\bar{q}\to {^1S_0^{~[1]}}] 
&=& 
\frac{32\,C_F}{3\,N_c^2}\,z\, \left(1-z\right)
\,,\nonumber \\
\cAnc[q\bar{q}\to {^1S_0^{~[8]}}] 
&=& 
\frac{32\,B_F}{3\,N_c^2}\,z\, \left(1-z\right)
\,,
\label{eq:cAnc1}
\end{eqnarray}
where $B_F=(N_c^2-4)/(4N_c)$ with $N_c=3$. Note that 
$\cAnc[gg\to {^1S_0^{[8]}}]$ is singular at $z=1$,
and regularized by the plus-prescription.
For spin triplet states one obtains
\begin{eqnarray}
\cAnc[gg\to {^3S_1^{[1]}}] 
&=&
\frac{256\,B_F}{6\,C_F\, N_C^2}\,
\frac{z}{(1-z)^2\,(1+z)^3}
\nonumber \\ &&\mbox{}
\times \bigg[
2+z+2z^2-4z^4-z^5
+2z^2(5+2z+z^2)\ln{z}
\bigg]
,\nonumber \\
\cAnc[gg\to {^3S_1^{\,[8]}}] 
&=&
\frac{1}{36 z (1-z)^2 (1+z)^3}
\bigg[
108+153 z+ 400 z^2+65 z^3-356 z^4-189 z^5
\nonumber \\ &&\mbox{}
-152 z^6-29 z^7
+\left(108 z+ 756 z^2+432 z^3+704z^4+260z^5
+76 z^6\right)\ln{z}
\bigg]
\,,\nonumber \\
\cAnc[gq \to {^3S_1^{\,[8]}}] 
&=&
\frac{T_F}{4}\,\left(1-z\right)\,\left(1+3 z\right)
+\frac{C_A}{4 C_F}\,\frac{1}{z}\bigg[(1-z)(2+z+2z^2)
\nonumber\\ &&\mbox{}
+2z\,(1+z)\,\ln{z}
\bigg]
\,,\nonumber \\
\cAnc[q\bar{q} \to {^3S_1^{\,[8]}}] 
&=&
-
\bigg[C_F\, (1-z)^2
+\frac{C_A}{3}\,(1+z+z^2)\bigg]
\left(\frac{1}{1-z}\right)_+
\, .
\label{eq:cAnc}
\end{eqnarray}
The function $\cAnc[q\bar{q} \to {^3S_1^{\,[8]}}]$
is also defined with the plus-prescription. 
The leading singular behavior of $\cAnc$ is given by
$\cAnc(z)\stackrel{z\rightarrow 1}{\sim} -C_A/(1-z)_+$ both  for 
$gg\rightarrow{^1S_0^{[8]}}$ and $q\bar{q}\rightarrow{^3S_1^{[8]}}$.
In the soft limit its behavior is insensitive to the details of the boundstate 
and only depends on its color configuration.

It is instructive to discuss the relation between the normalizations of
different processes leading to the same boundstate.
For instance, the normalization ${\cal N}_{ij\to T}$ (see Tab.~\ref{tb:XSnorm})
for the process $gq\rightarrow{^1S_0^{[1,8]}}X$ 
is fixed by $gg\rightarrow {^1S_0^{[1,8]}}$,
because in the collinear limit this cross section factorizes 
into the corresponding LO process and the $P_{gq}$ splitting function.
As a consequence the cancellation of the factorization scale dependence
happens among the $gg$ and $gq$ initiated reactions. Similarly, the
normalization of $gq\rightarrow{^3S_1^{[8]}}$ is fixed by 
$q\bar{q}\rightarrow{^3S_1^{[8]}}$. In contrast, 
the processes $q\bar{q}\rightarrow{^1S_0^{[1,8]}}$ and
$gg\rightarrow {^3S_1^{[1,8]}}$ are forbidden at LO,
hence the corrections have to be collinearly finite.
{In comparison to Ref.~\cite{Hagiwara:2008df} the
combinations ${\cal A}_c + {\cal A}_{nc}$ include terms that vanish in the
limit $z\to1$. Furthermore subprocesses that appear for the first time in
${\cal O}(\alpha_s^3)$ were neglected in Ref.~\cite{Hagiwara:2008df}. The
relative size of these terms will be adressed below.}

\begin{table}[t]
\begin{center}
{
\renewcommand{\arraystretch}{1.4}
\begin{tabular}{l|lll|lll}
\hline
 ~~~    &
\multicolumn{3}{c|}
            {${\cal L}\otimes F[ij\to T^{[1]}]\times 10^{\scalefont{0.7}6}$
              ~~[GeV$^{-2}$] } 
&
            \multicolumn{3}{c}
            {${\cal L}\otimes F[ij\to T^{[8]}]\times 10^{\scalefont{0.7}6}$
              ~~[GeV$^{-2}$] }
\\
\hline
\hline
$gg\rightarrow {^1}S_0^{[1,8]}$  &
$20.7$&$ 21.2$&$ 20.9$ &
$63.2$&$ 62.7$&$ 60.2$\\
\hline
$gq \rightarrow {^1}S_0^{[1,8]}$  &
$-0.795 $&$ -1.74 $&$ -2.19$ &
$-1.99 $&$ -4.36 $&$ -5.47$ \\
\hline
$q\bar{q} \rightarrow {^1}S_0^{[1,8]}$  &
$0.00664 $&$ 0.00509 $&$ 0.00398 $&
$0.0166 $&$ 0.0127 $&$ 0.00995 $\\
\hline
$gg\rightarrow {^3}S_1^{[1,8]}$  &
$ 0.175 $&$ 0.127  $&$ 0.0936  $ &
$ 6.06 $&$ 4.26  $&$ 3.07 $\\
\hline
$gq \rightarrow {^3}S_1^{[8]}$  &
&---&&
$3.99 $&$ 1.68 $&$ 0.279 $
\\
\hline
$q\bar{q} \rightarrow {^3}S_1^{[8]}$  &
&---&&
$ 23.1 $&$ 23.8 $&$ 23.6 $\\
\hline
 total: $\left({^1S_0}+{^3S_1}\right)^{[1,8]}$ &
 20.0 & 19.6 & 18.8 & 94.3 & 88.1 & 81.8 \\
\hline
\end{tabular}
\renewcommand{\arraystretch}{1.0}
}
\caption{
The convolution ${\cal L}\otimes F$ for LHC at the reference point $M=2m_t$, 
for the production of color
  singlet and octet states. The three columns correspond to the scale 
choices $\mur=\mu_f=(m_t,\, 2m_t,\, 4m_t)$.
}
\label{tb:LxF}
\end{center}
\end{table}

Let us now start the numerical analysis. The partonic cross 
sections have to be convoluted with the parton distribution 
functions (PDFs) in order to arrive at the hadronic cross section.
We use the CTEQ6.5~\cite{Tung:2006tb} set for the PDFs
and take $\alpha_s^{(5)}(M_Z)=0.118$,
$m_t=172.4$~GeV and $\sqrt{S}=14$~TeV as input values. The running of
$\alpha_s^{(5)}(\mur)$, which is the input for the partonic cross
sections, is evaluated with the help of {\tt
  RunDec}~\cite{Chetyrkin:2000yt},
using the four-loop approximation of the $\beta$ function.
This leads to $\alpha^{(5)}_s(\mur)=(0.1077, 0.09832, 0.09050)$
for $\mur=(m_t, 2m_t, 4m_t)$.
Furthermore we identify renormalization and factorization scales
 ($\mu_f=\mur$).

As stated above, the cross section factors into the convolution
${\cal L}\otimes F$ and the Green's function. To discuss the relative
importance of the various contributions individually the results 
for the subprocesses without the factor 
${\rm Im}\, G(M+i\Gamma_t)/m_t^2$ are given in
Tab.~\ref{tb:LxF}.
Note that color-singlet $t\bar{t}$ production is dominated by
by $gg\rightarrow {^1S_0^{[1]}}$.
Color-octet production is dominated by $gg\rightarrow{^1S_0^{[8]}}$ plus 
a $25\%$ contribution from $q\bar{q}\rightarrow {^3S_1^{[8]}}$.
{The size of the remaining subprocesses (neglected in
  Ref.~\cite{Hagiwara:2008df}) amounts to
  five to ten percent and is strongly scale dependent.}
The variation of $\mu$ (recall $\mu = \mu_f=\mur$) between $m_t$ and $4m_t$
leads to changes of ${\cal L}\otimes F$ by $\pm 3\%$ and {$\pm 7\%$}
for the total singlet and octet production, respectively.
In these channels the real radiation of partons 
contains large logarithmic contributions in the NLO corrections.
In combination with the rapidly varying parton luminosity 
{these logarithms} make up for a major part of the numbers quoted in
Tab.~\ref{tb:LxF}. The origin of these large logarithms can be traced to the 
singular behavior of the cross section near $z\approx 1$,
regularized by plus-distributions.
There exists well established technology for the resummation 
of these large logarithms to all orders in perturbation theory.
We will address this issue next.

\section{Soft gluon resummation}
\label{sec::resum}

The parton channels, which exhibit enhancement due to soft gluon emission are 
$gg\rightarrow {^1S_0^{[1]}}$, $gg\rightarrow {^1S_0^{[8]}}$, 
and $q\bar{q}\rightarrow {^3S_1^{[8]}}$
(see Eqs.~(\ref{eq:cAc}) and (\ref{eq:cAnc})).
The relevant logarithms are contained both in ${\cal A}_c$ (from 
initial state radiation) and 
${\cal A}_{nc}$ (from FSR) and read for the three leading processes:
\begin{eqnarray}
  {\cal A}_{thr log}
  \big[gg\rightarrow {^1S_0^{[1]}}\big]
  &=&
  4 C_A D_1
  - 2 C_A \ln\left(\frac{\mu_f^2}{M^2}\right)
     D_0
  -\frac{\beta_0}{2}\,\delta(1-z)\,\ln\left(\frac{\mu_f^2}{M^2}\right)\,,
\nonumber 
\\
  {\cal A}_{thr log}
  \big[gg\rightarrow {^1S_0^{[8]}}\big]
  &=&
  {\cal A}_{thr log}
  \big[gg\rightarrow {^1S_0^{[1]}}\big]
  - C_A D_0 \,,
\nonumber 
\\
  {\cal A}_{thr log}
  \big[q\bar{q}\rightarrow {^3S_1^{[8]}}\big]
  &=&
  4 C_F D_1
  - \left(2 C_F \ln\left(\frac{\mu_f^2}{M^2}\right)+C_A\right) D_0
  -\frac{3\,C_F}{2}\,\delta(1-z)\,\ln\left(\frac{\mu_f^2}{M^2}\right)\,,
  \label{eq:qqbarthrlog}
\end{eqnarray}
where $D_l = [\ln^{l}(1-z)/(1-z)]_+$ denote the plus-distributions and 
all $\ln\mu_f^2/M^2$ parts are included in the definition of threshold logarithm. 
Whether the threshold logarithms are enhanced or not depends on the behavior of the parton
luminosity functions near the kinematical end point $\tau=\rho$.
To investigate the size of the threshold logarithms,
we evaluate the contribution of the factorized hard scattering
contribution convoluted with the PDFs, i.e. ${\cal L}\otimes F$
separately for the three contributions:
tree-level, singular and regular terms.
(The hard corrections $(1+(\alpha_s/\pi)\, {\cal C})$ are common to all).
The threshold enhanced contributions are defined in Eq.~(\ref{eq:qqbarthrlog})
{and correspond exactly to the terms included in
  Ref.~\cite{Hagiwara:2008df}}, while regular
terms correspond to the remainder of $\cAc+\cAnc$ in
Eqs.~(\ref{eq:cAc}) and (\ref{eq:cAnc}) without plus distributions.
For $M=2m_t$ and $\sqrt{S}=14\,{\rm TeV}$ we obtain the following results
\begin{eqnarray}
\big({\cal L}\otimes F\big)\big[gg\to {^1S_0^{[1]}}\big]
&=&
\left\{
\begin{array}{l}
14.5+\big(4.53 + 1.68\big)_{\cal A} \\
14.0+\big(5.66 + 1.58\big)_{\cal A} \\
13.0+\big(6.37 + 1.48\big)_{\cal A} \\
\end{array}
\right\}
\times 10^{-6}~{\rm GeV^{-2}},
\nonumber \\
%
\big({\cal L}\otimes F\big)\big[gg\rightarrow {^1S_0^{[8]}}\big]
&=&
\left\{
\begin{array}{l}
39.3+\big(16.6 + 7.26\big)_{\cal A} \\
37.4+\big(18.8 + 6.52\big)_{\cal A} \\
34.4+\big(20.0 + 5.83\big)_{\cal A} \\
\end{array}
\right\}
\times 10^{-6}~{\rm GeV^{-2}},
\nonumber \\
%
\big({\cal L}\otimes F\big)\big[q\bar{q}\rightarrow {^3S_1^{[8]}}\big]
&=&
\left\{
\begin{array}{l}
16.7+\big(3.50 + 2.91\big)_{\cal A} \\
16.8+\big(3.41 + 3.56\big)_{\cal A} \\
16.4+\big(3.28 + 3.97\big)_{\cal A} \\
\end{array}
\right\}
\times 10^{-6}~{\rm GeV^{-2}}.
\label{eq:threshnumbers}
\end{eqnarray}
The three lines correspond to $\mu=\muf=\mur=(m_t,2m_t,4m_t)$.
We note that in all three cases the contribution of the 
threshold enhanced terms from Eq.~(\ref{eq:qqbarthrlog}) is large, 
although the regular terms in the case of $q\bar{q}\rightarrow {^3S_1^{[8]}}$
are of the same order. 
{Technically the matching applied in Ref.~\cite{Hagiwara:2008df} corresponds to
neglect all terms which vanish
exactly at threshold that is for $z=1$, i.e. Eqs.~(\ref{eq:cAnc1})
and~(\ref{eq:cAnc}) of Section~\ref{sec::hard}.
The regular terms in Eq.~(\ref{eq:threshnumbers}), which have not been
accounted for in the recent analysis of Ref.~\cite{Hagiwara:2008df},
are of the same order as the NLO sub-processes as given in Tab.~\ref{tb:LxF}.
}

Threshold resummation proceeds conveniently in Mellin-space. 
To that end we calculate the Mellin moments with respect to $z=M^2/\hat{s}$ 
according to 
\begin{eqnarray}
  \label{eq:mellindef}
 F^N_{ij\to T}(M^2,\mufs) 
&=&
  \int\limits_{0}^{1}\,dz\, z^{N-1}\,
  F_{ij\to T}(\hat s, M^2,\mufs)\, .
\end{eqnarray}

Then, the Mellin-space expression for the threshold enhanced terms listed in
Eq.~(\ref{eq:qqbarthrlog}) read (see also \cite{Cacciari:1999sy,Moch:2005ba})
\begin{eqnarray}
  \label{eq:res-gg1S01}
  {\cal A}^N_{thr log}
  \big[gg\rightarrow {^1S_0^{[1]}}\big] &=& 
         2 \* \ca \* \lnNs 
       + \ca \* \lnN \* \left( 4 \* \gE - 2 \* \Lqf \right) 
\nonumber\\
&&
       + \ca \* \left( 2 \* \z2 + 2 \* \gE^2 - 2  \* \gE \* \Lqf \right)
       + {1 \over 2} \* \beta_0 \* \Lqf
\, ,
\nonumber\\
  \label{eq:res-gg1S08}
  {\cal A}^N_{thr log}
  \big[gg\rightarrow {^1S_0^{[8]}}\big]
  &=&
  {\cal A}^N_{thr log}
  \big[gg\rightarrow {^1S_0^{[8]}}\big]
       + \ca \* \lnN 
       + \ca \* \gE 
\, ,
\nonumber\\
  \label{eq:res-qq3S18}
  {\cal A}^N_{thr log}
  \big[q\bar{q}\rightarrow {^3S_1^{[8]}}\big]
  &=&
         2 \* \cf \* \lnNs 
       + \cf \* \lnN \* \left( 4\*\gE - 2  \* \Lqf \right)
       + \ca \* \lnN
\nonumber\\
&&
       + \cf \* \left( 2 \* \z2 + 2 \* \gE^2 + {3 \over 2} \* \Lqf 
         - 2  \* \gE \* \Lqf \right)
       + \ca \* \gE
\, ,
\end{eqnarray}
where we have kept all dominant terms in the large-$N$ limit and neglected
power suppressed terms of order $1/N$.
$\gE$ is the Euler-Mascheroni constant ($\gE =  0.577215\ldots$).

The resummed expressions (defined in the \MSbar-scheme) 
for the individual color structures of the hard cross sections $F$ of Eq.~(\ref{eq:F}) 
are given by a single exponential in Mellin-space (see e.g. Refs.~\cite{Moch:2005ba,Contopanagos:1996nh,Catani:1996yz})
\begin{equation}
\label{eq:sigmaNres}
{F^N_{ij \to T}(M^2,\mufs) \over F^{(0), N}_{ij \to T}(M^2,\mufs)} = 
  g^0_{ij \to T}(\mts,\mufs,\murs) \, \Delta^{N+1}_{ij \to T}(\mts,\mufs,\murs) + 
  {\cal O}(N^{-1}\ln^n N) \, ,
\end{equation}
where $F^{(0), N}_{ij \to T}$ denotes the tree level term in Eq.~(\ref{eq:F})
and the  
exponents are commonly expressed as 
\begin{equation}
\label{eq:GNexp}
  \ln \Delta^N_{ij \to T}  = 
  \ln N \cdot g^1_{ij}(\lambda)  +  g^2_{ij \to T}(\lambda)  + \dots\, ,
\end{equation}
where $\lambda = \beta_0\, \alpha_s\, \ln N/(4 \pi)$. 
To next-to-leading logarithmic (NLL) accuracy the (universal) functions
$g^1_{ij}$ as well as the functions $g^{2}_{ij \to T}$  
are relevant in Eq.~(\ref{eq:GNexp}), see Ref.~\cite{Moch:2008qy} for the
extension to NNLL accuracy. Explicit expressions are
\begin{eqnarray}
  \label{eq:g1res}
  g^1_{q{\bar q}} &=& 
          {\Aq1 \over \b0} \*  [
            2
          - 2\*\ln(1-2\*\lambda)
          + \lambda^{-1} \* \ln(1-2\*\lambda) 
          ]
\, ,
\nonumber\\
  \label{eq:g2res}
  g^2_{q{\bar q} \to T[1]} &=& 
        \biggl(
            {\Aq1 \* \beta_1 \over \b0^3}
          - {\Aq2 \over \b0^2}
        \biggr) \* [
            2\*\lambda
          + \ln(1-2\*\lambda)
          ]
          + {\Aq1 \* \b1 \over 2 \* \b0^3} \* \ln^2(1-2\*\lambda)
\nonumber\\
&&\mbox{}
       - 2 \* {\Aq1 \over \b0} \* \gE \* \ln(1-2\*\lambda)
       + \Lqr \* {\Aq1 \over \b0} \* \ln(1-2\*\lambda)
       + 2\* \Lfr \* {\Aq1 \over \b0} \* \lambda
\, ,
\nonumber\\
  g^2_{q{\bar q} \to T[8]} &=& 
       g^2_{q{\bar q} \to T[1]} 
          - {\DQQ1 \over 2 \* \b0} \* \ln(1 - 2 \* \lambda)
\, ,
\end{eqnarray}
%
where the full dependence on $\mur$ and $\muf$ has been kept. 
The gluonic expressions $g^1_{gg}$ and $g^2_{gg \to T}$ 
are obtained with the obvious replacement $A^{(i)}_q \to A^{(i)}_g$.
The perturbative expansions of the anomalous dimensions 
{are universal and well-known}. We have~\cite{Kodaira:1981nh}
\begin{eqnarray}
\label{eq:Aqexp}
  A^{(1)}_q & = & 4\, C_F 
\, ,
\nonumber \\
  A^{(2)}_q & = & 8\, C_F \left[ \left( \frac{67}{18}
     - \zeta_2 \right) C_A - \frac{5}{9}\,n_f \right] 
\, ,
\nonumber \\
  \label{eq:Dqqbarexp}
  D^{(1)}_{Q{\bar Q}} & = & 4\, C_A 
\, ,
\end{eqnarray}
and all gluonic quantities are given by multiplying $A^{(i)}_q$ by $C_A/C_F$. 
We also give explicit results for the matching functions $g^0_{ij \to T}$ in Eq.~(\ref{eq:sigmaNres}),
\begin{eqnarray}
  \label{eq:g0gg1}
  g^0_{gg \to T[1]} &=&
  1 + {\alpha_s \over \pi} \*
          \left\{
         \ca \* \left[ 2 \* \z2 + 2 \* \gE^2 - 2  \* \gE \* \Lqf \right]
       + {1 \over 2} \* \beta_0 \* \Lqf
          \right\}
\, ,
\nonumber\\
  \label{eq:g0res}
  g^0_{gg \to T[8]} &=& g^0_{gg \to T[1]} + 
        {\alpha_s \over \pi} \* \ca \* \gE 
\, ,
\nonumber\\
  \label{eq:g0qq8}
  g^0_{q{\bar q} \to T[8]} &=&
  1 + {\alpha_s \over \pi} \*
          \left\{
         \cf \* \left[ 2 \* \z2 + 2 \* \gE^2 + {3 \over 2} \* \Lqf - 2  \* \gE
           \* \Lqf \right]
       + \ca \* \gE
          \right\}
\, .
\end{eqnarray}

For phenomenological applications~\cite{Bonciani:1998vc,Bonciani:1998vcerr} of
soft-gluon resummation  
at the parton level one introduces an improved (resummed) hard cross section $F^{\rm res}$, 
which is obtained by an inverse Mellin transformation as follows,
\begin{eqnarray}
\label{eq:defsigmares}
F^{\rm res}_{ij\to T}(\hat s,M^2,\mufs) &=& 
\int\limits_{c-{\rm i}\infty}^{c+{\rm i}\infty}\,
{dN \over 2\pi {\rm i}}\, x^{-N}\,
\left(
  F^N_{ij \to T}(M^2,\mufs) - 
  \left. F^N_{ij \to T}(M^2,\mufs) \right|_{\rm NLO} 
\right)
\nonumber\\ &&
  + 
  F^{\rm NLO}_{ij \to T}(\hat s,M^2,\mufs)\, .
\end{eqnarray}
Here $F^{\rm NLO}_{ij \to T}$ is the standard fixed order 
cross section at NLO in QCD, while $F^N_{ij \to T}\bigr|_{\rm NLO}$ 
is the perturbative truncation at the same order in $\alpha_s$ obtained by
employing Eq.~(\ref{eq:res-qq3S18}). 
That is to say that for the matching we have fully expanded all formulae 
consistently to ${\cal O}(\alpha_s)$. 
This adds the hard coefficients $\cCh$  of Eq.~(\ref{eq:cCh}) 
to the results Eqs.~(\ref{eq:res-qq3S18}) and (\ref{eq:g0gg1}). 
In this way, the right-hand side of Eq.~(\ref{eq:defsigmares}) reproduces the
fixed order results and resums soft-gluon effects beyond NLO to NLL accuracy.

In Section~\ref{sec::results} we employ Eq.~(\ref{eq:defsigmares}) for
phenomenological predictions by performing the inverse Mellin transform numerically.
To that end, one should note that the treatment of the precise numerical
matching to the exact NLO hard cross section is a matter of choice
since different schemes lead only to differences which are
formally of higher order.
We have found that the application of the resummed result is well justified 
when the kinetic energy of the top-quark pair is a few GeV {or less}, see
e.g. Ref.~\cite{Moch:2008qy}, 
where the precise numerical value is not important. 
Another issue concerns the constant terms in Eq.~(\ref{eq:g0qq8}) 
which are sometimes modified to include formally sub-leading 
(but numerically not insignificant) terms, see for instance
Ref.~\cite{Bonciani:1998vc,Bonciani:1998vcerr}. 
As just explained, in the present analysis we adopt the minimal approach, 
i.e. we apply Eq.~(\ref{eq:g0qq8}) (including the hard coefficients $\cCh$ of
Eq.~(\ref{eq:cCh}))  
and account for all regular terms in Eq.~(\ref{eq:threshnumbers}) through
matching to NLO. 

In Tab.~\ref{tab:nlo-res} we compare the fixed-order NLO and resumed result 
of the convolution ${\cal L}\otimes F$. One observes an enhancement up to
about 10\% depending on the process.

\begin{table}[t]
\begin{center}
\renewcommand{\arraystretch}{1.4}
\begin{tabular}{c|c|c|c|c|c|c}
\hline
&
\multicolumn{3}{|c|}{NLO }&
\multicolumn{3}{c}{resummed}\\ \hline\hline
$gg \to ^1S_0^{[1]}$ &
20.7 & 21.2 & 20.9
& 22.0 & 23.2 & 24.0 \\
$gg \to ^1S_0^{[8]}$ &
63.2 & 62.7 & 60.2
& 67.8 & 69.7 & 70.6 \\
$q\bar q \to ^3S_1^{[8]}$ &
23.1 & 23.8 & 23.6
& 23.8 & 24.0 & 23.6 \\
\hline
\end{tabular}
\caption{\label{tab:nlo-res}Comparison of the NLO and resummed 
  result of the convolution ${\cal L}\otimes F$ (in $10^{-6}$~GeV$^{-2}$)
  for LHC at the reference point $M=2m_t$.
  The three columns correspond to the scale
  choices $\mur=\mu_f=(m_t,\, 2m_t,\, 4m_t)$.
  The NLO results can also be found in Tab.~\ref{tb:LxF}.
}
\renewcommand{\arraystretch}{1.}
\end{center}
\end{table}


\section{Boundstate corrections}
\label{sec::bound}

Let us next discuss the boundstate corrections.
As mentioned above, the convolution of $F_{ij\to T}$ with the parton luminosities
provides the normalization of the differential cross section, while its shape
is mainly determined by the non-relativistic Green's function.
The latter describes the long-distance
evolution of the top quark pair produced near threshold.
The kinematics of the produced top quark pair is nonrelativistic, and the
dynamics is governed by exchange of potential gluons leading to the
formation of quasi-boundstates.
The corresponding potential is given at NLO by 
\begin{eqnarray}
  \widetilde{V}_C^{[1,8]}({\vec q})
  &=&
  -\frac{ 4\pi\alpha_s(\mur)\, C^{[1,8]}}{ {\vec q}\,^2}\,
  \bigg[1+\frac{\alpha_s(\mur)}{4\pi}
    \bigg( \beta_0 \, \ln\frac{\murs}{{\vec q}\,^2} +a_1
    \bigg)
    \bigg]\,,
    \label{eq::VQCD}
\end{eqnarray}
with $C^{[1]} =C_F=4/3$ and
$C^{[8]} =C_F-C_A/2=-1/6$, and $a_1=(31/9)\,C_A-(20/9)\,T_F\,n_f$.

\begin{figure}[t]
  \begin{center}
    \includegraphics[width=0.8\textwidth]{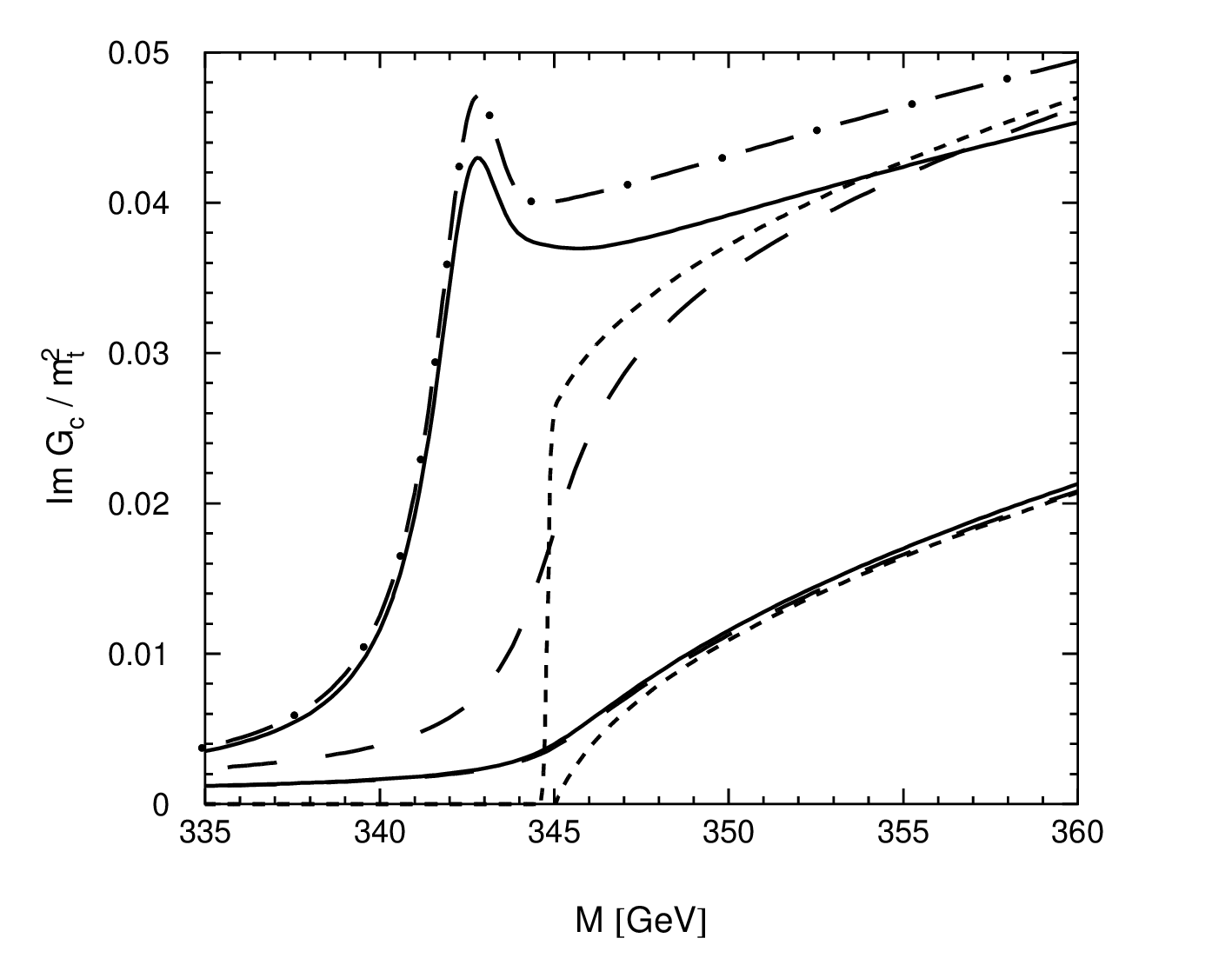}
    \caption{\label{fig:GF}
      Imaginary part of the Green's functions for the color singlet 
      (upper solid  line) and color octet (lower solid line) cases 
      as functions of top quark invariant mass.
      For comparison, also the expansions of $G$ in fixed order up
      to ${\cal O}(\alpha_s)$ with (dashed) and without 
      (dotted line) $\Gamma_t$ are plotted. The imaginary part of the 
      NNLO Green's function for the color-singlet case is shown as dash-dotted
      line.}
  \end{center}
\end{figure}

The color-singlet Green's function feels an attractive force, the
color-octet Green's function is governed by repulsion and
thus does not develop a boundstate.
They are both defined as the solutions of the Schr\"odinger equations
\begin{eqnarray}
  \left\{2m_t
  +\left[\frac{(-i\,\nabla)^2}{m_t} + V_C^{[1,8]}(\vec{r}\,)\right]
  - (M+i\Gamma_t)\right\} \,
  G^{[1,8]}({\vec r}; M+i\Gamma_t )
  &=&\delta^{(3)} ({\vec r}\,)\,.
  \label{eq:defG}
\end{eqnarray}
For the Green's function at zero-distance, 
the NLO result is known 
in a compact form~\cite{Beneke:1999qg} (see also~\cite{Pineda:2006ri}) 
\begin{eqnarray}
G^{[1,8]}(M+i\Gamma_t)
&\equiv&
G^{[1,8]}(\vec{r}=0;M+i\Gamma_t)
=
\frac{C^{[1,8]}\alpha_s(\mur)\, m_t^2}{4\pi}\,
\left[g_{\rm LO}+ \frac{\alpha_s(\mur)}{4\pi}\, g_{\rm NLO}+\cdots\right],
\nonumber\\
g_{LO}&=&
-\frac{1}{2\kappa}+L-\psi^{(0)},
\nonumber\\
g_{NLO}&=&
\beta_0
\bigg[
L^2
-2 L\,
\bigg(
\psi^{(0)}-\kappa \,\psi^{(1)}
\bigg)
+\kappa\,\psi^{(2)}
+\big(\psi^{(0)}\big)^2
-3\,\psi^{(1)}
-2\,\kappa\,\psi^{(0)}\,\psi^{(1)}
\nonumber \\ &&\mbox{} 
+ 4\,\,\, {_{4}F_3\bigg(1,1,1,1;2,2,1-\kappa;1\bigg)}
\bigg]
+ a_1
\bigg[
L-\psi^{(0)}+\kappa\,\psi^{(1)}
\bigg],
\label{eq::GCana}
\end{eqnarray}
with
\begin{eqnarray}
  \kappa
  \equiv
  \frac{i\,C^{[1,8]}\,\alpha_s(\mur)}{2\,\varv}, ~~~ 
  \varv=\sqrt{\frac{M+i\Gamma_t-2m_t}{m_t}}.
  \label{eq:kapp}
\end{eqnarray}
Here $L=\ln\left(i\mur/\left(2m_t\,\varv\right)\right)$
and $\psi^{(n)}=\psi^{(n)}(1-\kappa)$ is the $n$-th derivative 
of $\psi(z)\equiv\gamma_E+(d/dz)\ln\Gamma(z)$ with argument $(1-\kappa)$.
The Green's function in Eq.~(\ref{eq::GCana}) correctly reproduces all the 
NLO terms in NRQCD, however, it is not sufficient to describe 
the behavior of the Green's function in the vicinity of boundstate 
poles. It is because the exact solution to the Schr\"odinger equation 
has only single poles in the boundstate energy 
$G^{[1]}\sim |\Psi(0)|^2/(M_n - M-i\Gamma_t)$, while Eq.~(\ref{eq::GCana})
is  an expansion around the LO boundstate poles and thus has 
multiple poles of a form $G\sim |\Psi_n^{(0)}(0)|^2/(M_n^{(0)}-M)^k$
($k=1,2$ at the NLO).
However, resummation of this multiple poles into single poles is
straightforward and well-known. 
We refer to Ref.~\cite{Beneke:1999qg} for further details.

In Fig.~\ref{fig:GF} we show the imaginary parts of the
color singlet and color octet Green's functions
in the threshold region. 
As input we use $m_t^{\rm PS}=170.1$~GeV,
which to NLO accuracy corresponds to $m_t=172.4\,{\rm GeV}$ \cite{ICHEP:2008},
and $\Gamma_t=1.36$~{\rm
  GeV}~\cite{Jezabek:1993wk,Czarnecki:1998qc,Chetyrkin:1999ju}.
At NLO the Green's function is separately renormalization scale
invariant and we are free to chose $\mur$ independent from
the hard process. A well-motivated physical
scale is $\mu_s = m_t C_F \alpha_s(\mu_s)=32.21$~GeV which
corresponds to twice the inverse Bohr radius. The corresponding 
$\alpha_s$ value used in Fig.~\ref{fig:GF} is 
$\alpha_s^{(n_f=5)}(\mu_s)=0.1401$.
It has been observed that the color-singlet Coulomb Green's function 
has a well-convergent perturbative series for this scale
choice~\cite{Beneke:2005hg}.  

In order to see the effect of Coulomb resummation, we plot for both 
color states three lines:
the full Green's function (solid line) and
the expansion of $G$ in fixed order up to ${\cal O}(\alpha_s)$ 
with and without top quark width
(dashed/dotted). 
The upper three lines in Fig.~\ref{fig:GF} correspond to the
color singlet case and the lower three to the color octet one.
The color-singlet Green's function shows a pronounced
peak which corresponds to the $t\bar{t}$ resonance below $2m_t$,
while for color octet there is no enhancement.
Note that the curve for the full octet Green's function is very close 
to the one-loop expansion (taking into account the finite
top quark width). Thus for the color octet state the Coulomb 
resummation effect is negligible. 
In addition, one more line (dash-dotted) 
for the color-singlet Green's function is plotted including the NNLO 
Coulomb potential, which is useful to estimate yet 
unknown boundstate corrections to the NLO 
color-singlet Green's function. 
As input value we again adopt the PS top quark mass~\cite{Beneke:1998rk} 
given above. Note that in the absence of 
full NNLO result for the Green's function and hard
correction, this improved Green's function would not be sufficient 
for a full NNLO prediction. Nevertheless, the difference between solid
and dash-dotted curves gives an indication of the intrinsic
uncertainties of the Green's function, which is roughly $10\%$.

The expansion of 
$G$ up to ${\cal O}(\alpha_s)$ is obtained 
from the $g_{LO}$ in Eq.~(\ref{eq::GCana}) as
\begin{eqnarray}
\frac{1}{m_t^2}\, {\rm Im} \,G_c &=&{\rm Im}\,
\bigg[ \frac{\varv}{4\pi}\,
\left(i+\frac{\alpha_s C^{[1,8]}}{\varv}\left[\frac{i\pi}{2}-\ln{\varv}\right] 
\right)
\bigg]+{\cal O}(\alpha_s^2). 
\end{eqnarray}
In the zero-width limit ($i\Gamma_t\rightarrow +i\,0$), 
the color-singlet curve for the expansion 
exhibits a step of height $\alpha_s\,C_F/8$ (for $M\to2m_t$), and the 
color-octet curve formally becomes negative for 
$\varv \leq -\alpha_s C^{[8]}\pi/2$ which corresponds to $M-2m_t < 0.23$~{\rm GeV}. 
Both for the singlet and octet case the fixed order  
result without $\Gamma_t$ the imaginary part of the Green's function 
vanishes below $2m_t$. 
The qualitative difference between the solid and the short-dashed curves
will be reflected in the comparison of our final results for the 
invariant mass distribution with the prediction based on a fixed order
calculation: {for the color-singlet curve we observe a sizable 
excess in the region below the nominal threshold up 
to roughly $5~{\rm GeV}$ above}. In the color-octet case, as a consequence
of the relative smallness of $C^{[8]}$, the prediction follows roughly
the Born approximation. Although the color-octet Green's function is 
significantly smaller than the singlet one, the relatively large hard
scattering factor ${\cal L}\otimes F$ for ${^1S_0^{[8]}}$ plus
  ${^3S_1^{[8]}}$, which exceeds the one for the singlet case by roughly
    a factor four, quickly over-compensates the effect of the Green's functions.

In the present paper we use the analytical result of the Green's function, which
includes the $\alpha_s$ correction (i.e. the second term in the square brackets
of Eq.~(\ref{eq::VQCD})) by means of the Rayleigh-Schr\"odinger perturbation
approach. 
In Ref.~\cite{Hagiwara:2008df} a numerical solution to Eq.~(\ref{eq:defG})
has been employed, which resums the $\alpha_s$ corrections to all order.
The numerical solution is more stable against scale variation and applicable
over a wide range of $\mur$. However, the difference between the two
approaches is below $2\%$ and formally of higher order. Extensive studies on
higher order effect to the color singlet Green's function exist in the
literature (see, e.g., Refs.~\cite{Hoang:2000yr,Beneke:2005hg}), including
different implementations of the Green's function. From the experience collected
in the linear collider studies on $t\bar{t}$ 
production, we expect rather large corrections from the variation of $\mur$ 
for the color singlet Green's function of about $20\%$ 
which is significantly bigger than the
estimate from the NNLO Green's function mentioned above. 
In contrast to the color-singlet case the higher order corrections to
the color octet Green's function are expected to be unimportant 
since there is no resonance enhancement and the color coefficient
$C^{[8]}$ is small.


\section{Invariant mass distribution}
\label{sec::results}

\begin{figure}[t]
  \begin{center}
    \includegraphics[width=0.8\textwidth]{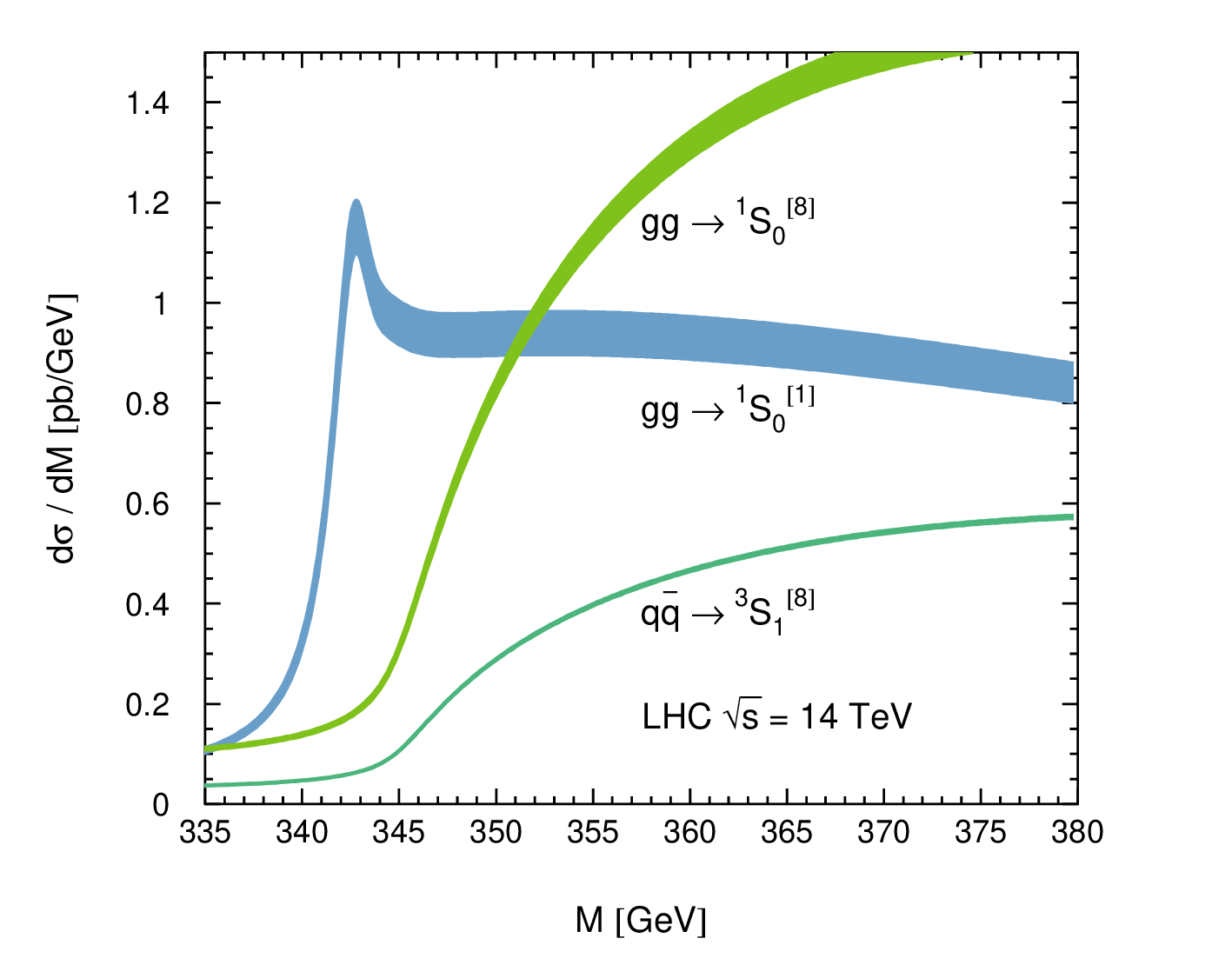}
    \caption{\label{fig::ind1}Invariant mass distributions for leading
      subprocesses:
      $gg\rightarrow {^1S_0^{[1,8]}}$ (blue and light green, respectively) and
      $q\bar{q}\rightarrow {^3S_1^{[8]}}$(green). For each process the
      bands take into account scale variation of the hard cross
      sections.
    }
    \label{fig:xs1}
  \end{center}
\end{figure}
\begin{figure}[t]
\begin{center}
\includegraphics[width=0.8\textwidth]{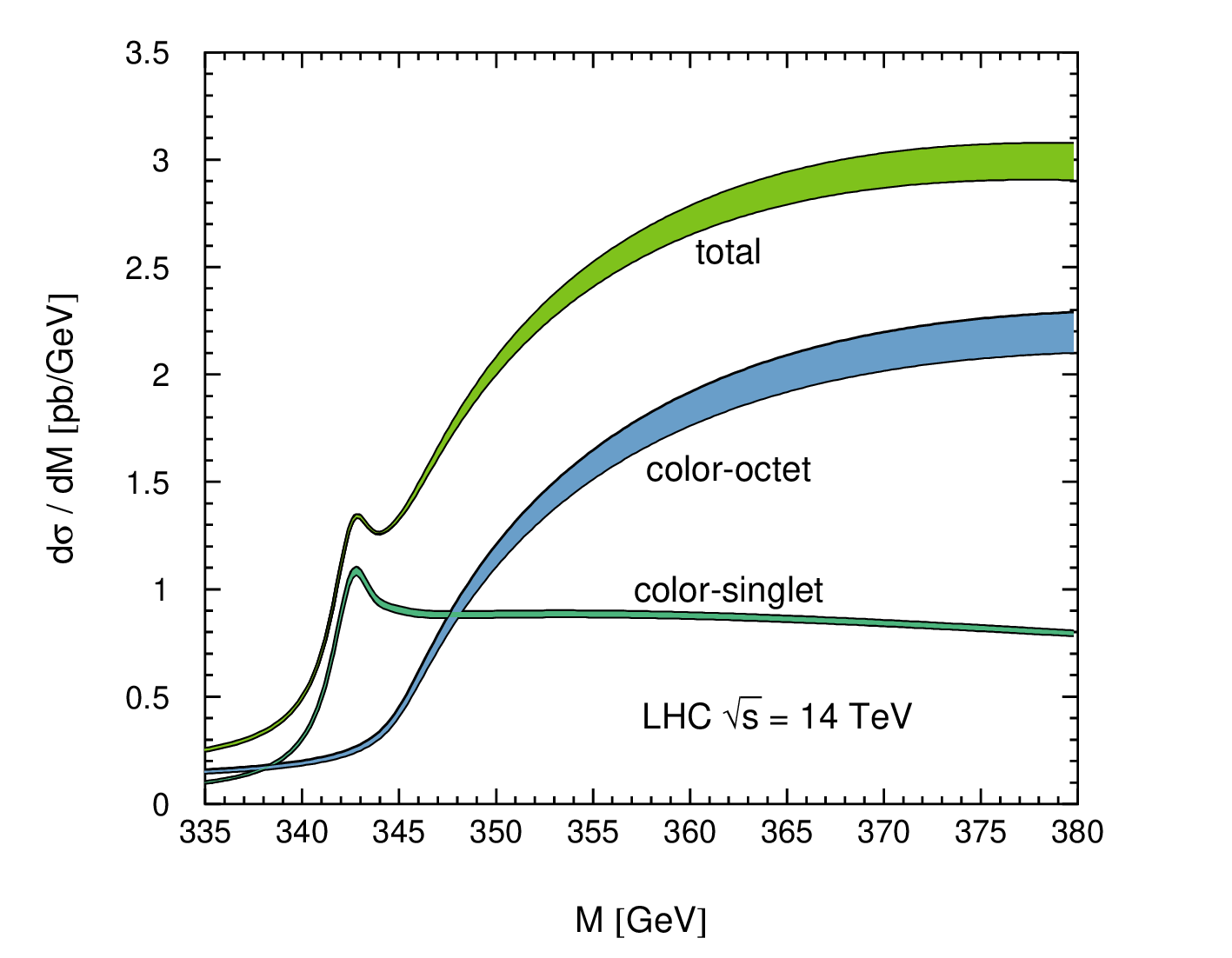}
\caption{
\label{fig:xs2}
Invariant mass distribution including all production 
channels shown in Tab.~\ref{tb:LxF}. The 
width of the bands reflect the scale dependence of the hard
scattering parts.
}
\end{center}
\end{figure}

We are now in the position to combine the results of the
preceeding Sections and discuss the cross section for the
invariant top quark distribution.

In Fig.~\ref{fig::ind1} the invariant mass distributions for
LHC ($\sqrt{S}=14$ TeV) is shown for the three dominant
processes. The bands reflect the scale
variation of the convolution ${\cal L}\otimes F$ which for
the color singlet case amounts to roughly $\pm 1\%$.
{The reduction as compared to Tab.~\ref{tb:LxF} and Fig.~\ref{fig:xs1}
  is due to a 
  compensation of the $\mu$ dependence after including the sub-leading
  NLO processes.
}
Note, however, that the corresponding Green's function
shows an uncertainty due to the renormalization scale variation of
about $20 \%$ which is well-known from top quark production
studies in $e^+ e^-$ collisions, consistent with the difference between
solid and dash-dotted curves in Fig.~\ref{fig:GF} and thus not discussed in
Fig.~\ref{fig::ind1}. This pattern is also evident from
Fig.~\ref{fig:xs2}, where all production channels as listed in
Tab.~\ref{tb:LxF} are included. The width of the bands is obtained from
varying
renormalization and factorization scales in the hard cross section
as described above. The additional uncertainty from the
Green's function, which we estimate 20\%
for the singlet and {below 5\%} for the octet case, is not included.

As expected, for $M<2 m_t$ the
production of $t{\bar{t}}$ pairs is dominated by the singlet
contribution. However, for $M>2 m_t$ one observes a strong raise
of the octet contributions, in particular of gluon induced 
subprocess which for $M\gsim 2m_t+5$~GeV becomes even 
larger than the corresponding singlet contribution.
For the color-octet case the scale dependence of the hard 
scattering amounts to $\pm 7\%$. 
Considering the threshold 
behavior as shown in Figs.~\ref{fig:xs1} and \ref{fig:xs2} 
it is clear, that the location of the
threshold is entirely governed by the behavior of the color singlet
($S$-wave) contribution. Thus, as a matter of principle, determining
the location of this step experimentally would allow for a top quark
mass measurement, which is conceptually very different from the one
based on the reconstruction of a (colored) single quark in the decay
chain $t\to Wb$. In fact, much of the detailed investigations of
$t\bar t$ threshold production at a linear collider were performed
for this particular relations between the location of the
color singlet quasi-boundstate pole of $t\bar{t}$ and the top 
quark $\overline{\mbox{MS}}$-mass.
{
The absolute normalization of the cross section is also sensitive
towards electroweak
corrections~\cite{Beenakker:1993yr,Kuhn:2005it,Bernreuther:2005ej,Kuhn:2006vh,Bernreuther:2006vg}
which are of the order of 5\%
close to threshold. For example, the difference between corrections
from a light ($M_h = 120$ GeV) and a heavy ($M_h = 1000$ GeV) Higgs
boson amounts to roughly 6\%~\cite{Kuhn:2006vh}.
}

\begin{figure}[t]
\begin{center}
\includegraphics[width=0.8\textwidth]{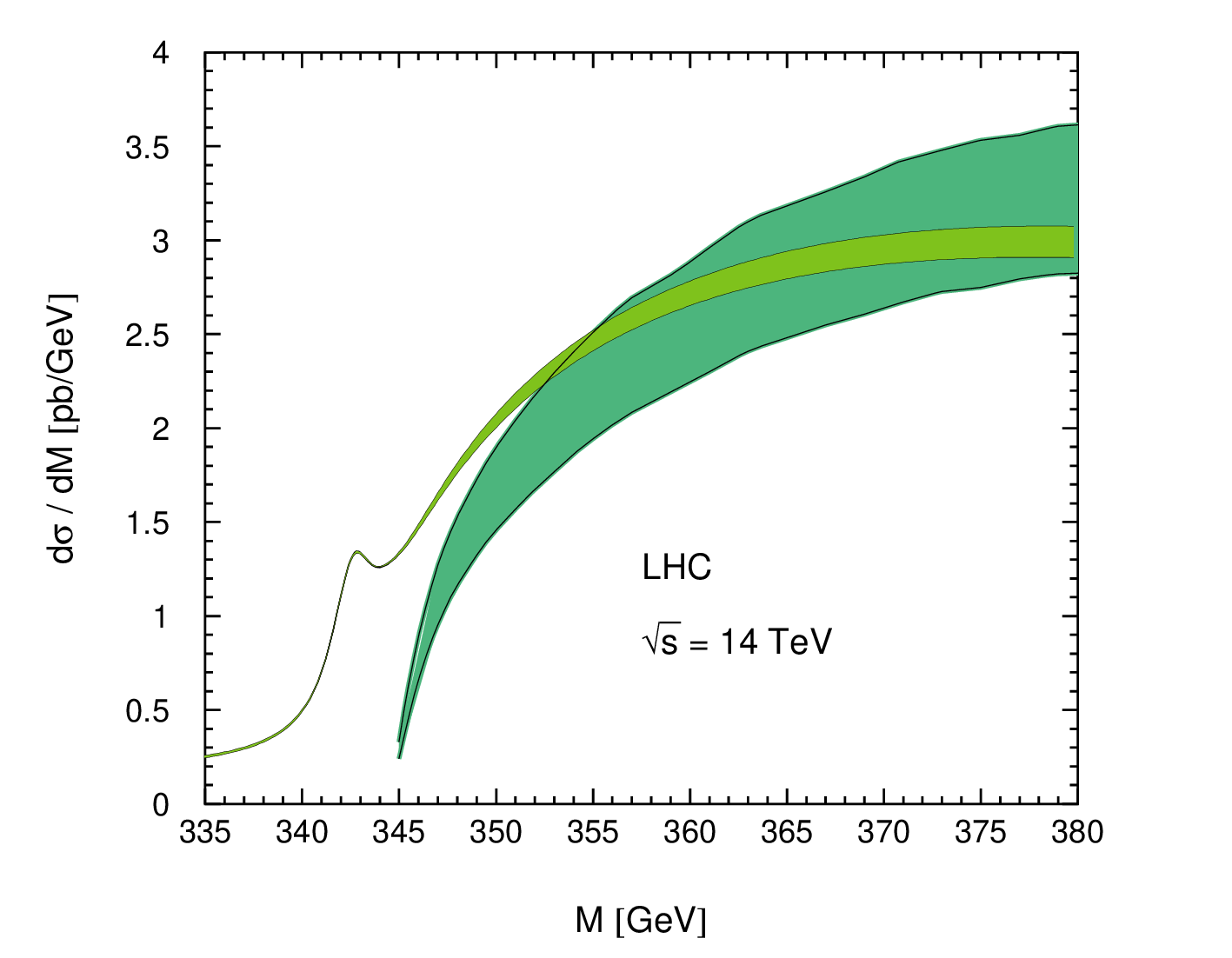}
\caption{
\label{fig:mtt_threshold}
Invariant mass distribution ${\rm d}\sigma/{\rm d}M$ from NRQCD
and for a fixed NLO for LHC with $\sqrt{s}=14~{\rm TeV}$. 
{The bands are due to scale variation from
$m_t$ to $4m_t$. For the NRQCD prediction the additional
uncertainty due to the Green's function estimated to 20\% (5\%)
for the colour singlet (octet) contribution is not included.}
}
\end{center}
\end{figure}
In Fig.~\ref{fig:mtt_threshold} the prediction for ${\rm d}\sigma/{\rm d}M$ 
based on NRQCD is compared with the one obtained from a fixed order NLO 
calculation for stable top quarks which is
obtained using the program HVQMNR \cite{Mangano:1991jk}.
As expected from the comparison of solid and dotted
curves in Fig.~\ref{fig:GF}, the two predictions overlap for 
invariant masses around $355~{\rm GeV}$. Above $355~{\rm GeV}$
relativistic corrections start to become important. From this comparison
we find an additional contribution to the total cross section for $t\bar{t}$
production of roughly 10~pb, which could become of relevance for 
precision measurements.
Note that the band of the NRQCD-based prediction only contains the uncertainty
from the scale variation of ${\cal L}\otimes F$ whereas the one of the Green's
function (which can reach up to 20\%, see Section~\ref{sec::bound}) is not
shown.

\begin{figure}[t]
\begin{center}
\includegraphics[width=0.8\textwidth]{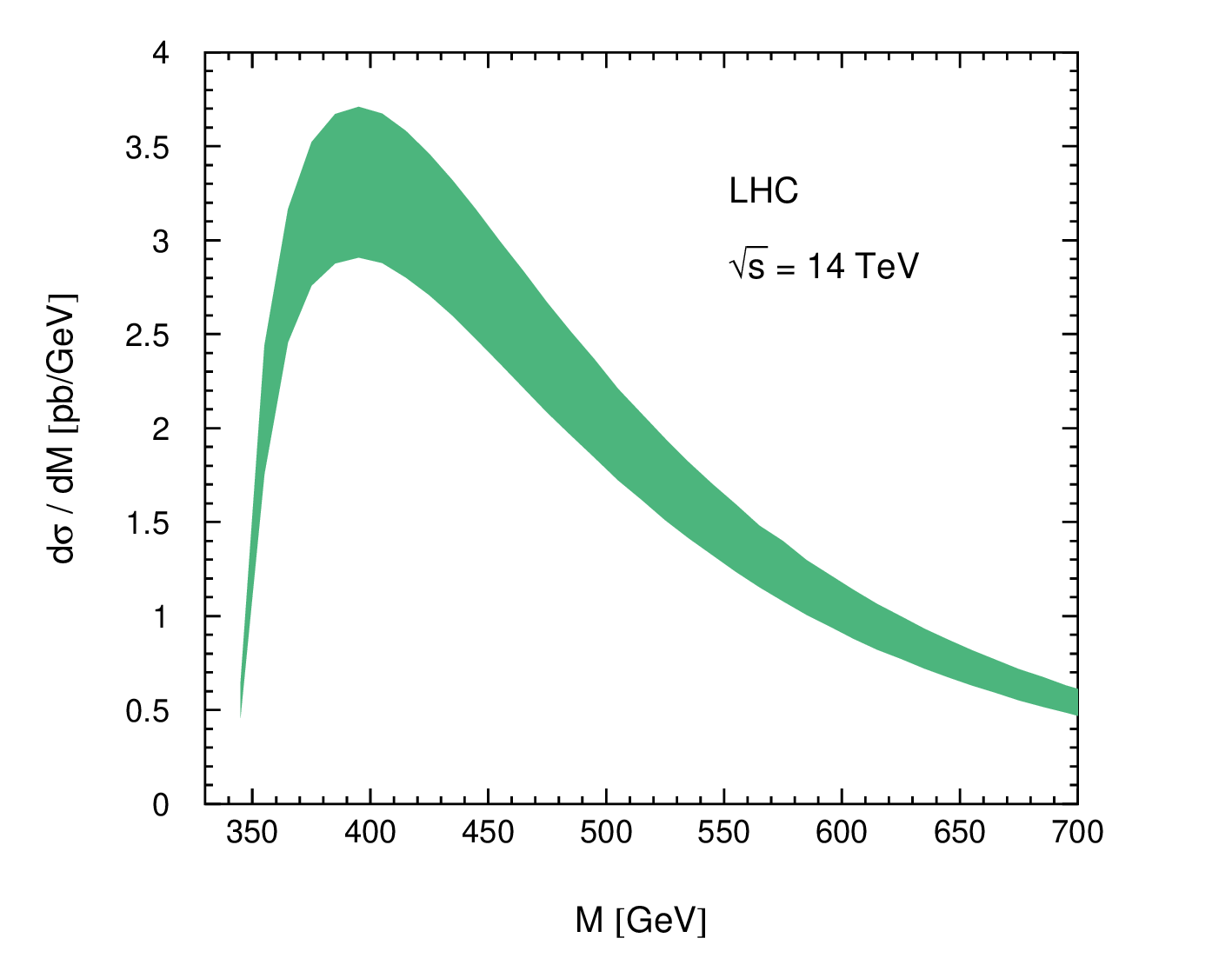}
\caption{
\label{fig:mtt_all}
Invariant mass distribution ${\rm d}\sigma/{\rm d}M$ from NLO calculation 
for LHC with $\sqrt{s}=14~{\rm TeV}$.
}
\end{center}
\end{figure}
The analysis of this work has concentrated on the threshold region and
is applicable for $M$ up $360~{\rm GeV}$ 
at most. {However, it is obvious, that the overall shape of 
${\rm d}\sigma/{\rm d} M$ will be distorted and the mean $\langle M\rangle$
shifted to smaller values, which might affect the global fit of 
${\rm d}\sigma/{\rm d} M$. 
}
In Fig.~\ref{fig:mtt_all} we present for comparison the NLO prediction
for ${\rm d}\sigma/{\rm d}M$ in the wide range up to $700~{\rm GeV}$.
The distribution reaches quickly its maximum of $3.3~{\rm pb/GeV}$
at around  $390~{\rm GeV}$ and then falls off slowly. It is remarkable
that its value at $370~{\rm GeV}$ is already not too far from the maximum
of the curve {and the threshold modifications thus affect a sizeable part of
the distribution}.

\begin{figure}[t]
\begin{center}
\includegraphics[width=0.8\textwidth]{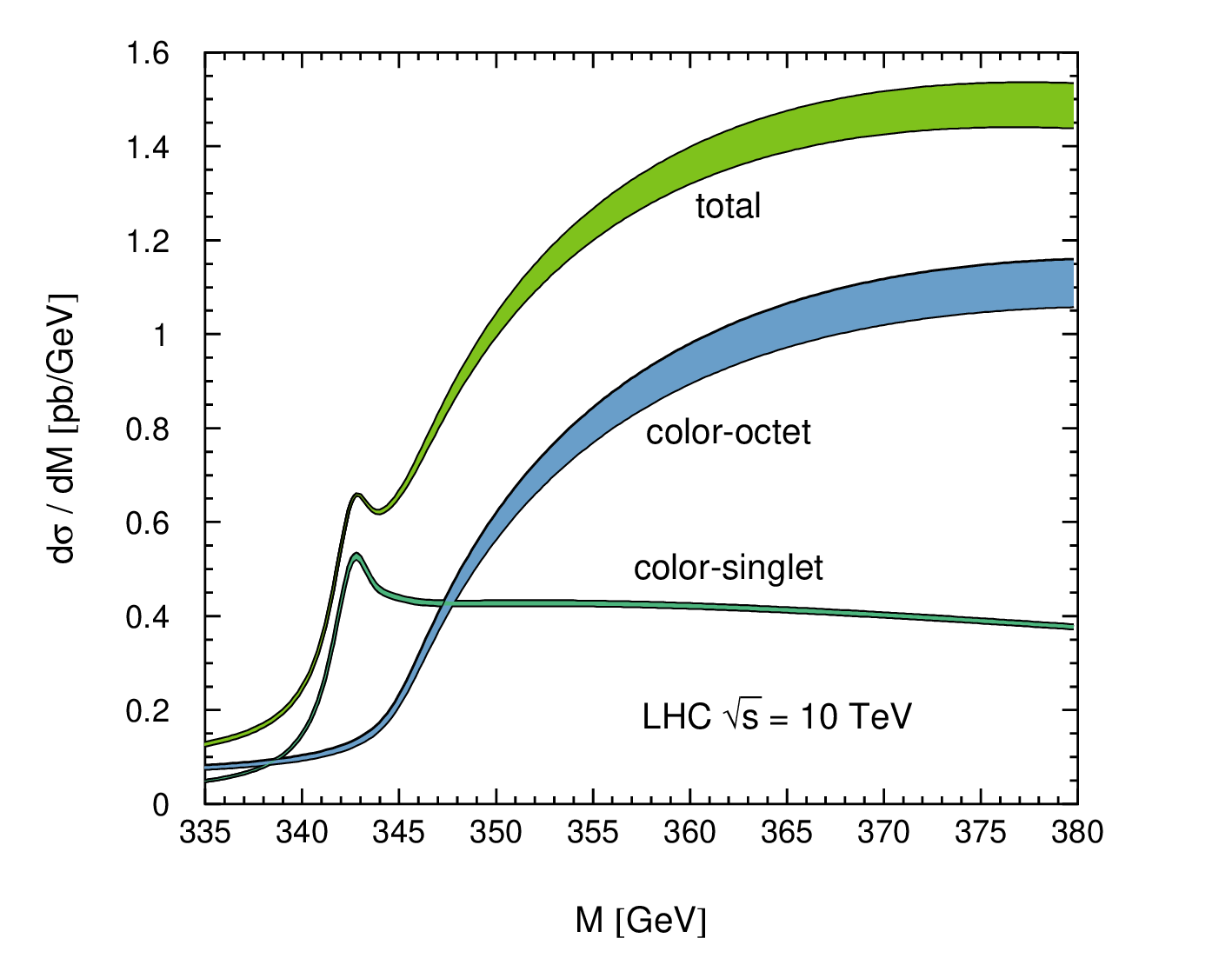}
\caption{
\label{fig:xs2lhc10}
Invariant mass distribution ${\rm d}\sigma/{\rm d}M$ for LHC with 
$\sqrt{s}=10{\rm TeV}$.
}
\end{center}
\end{figure}

\begin{figure}[t]
\begin{center}
\includegraphics[width=0.8\textwidth]{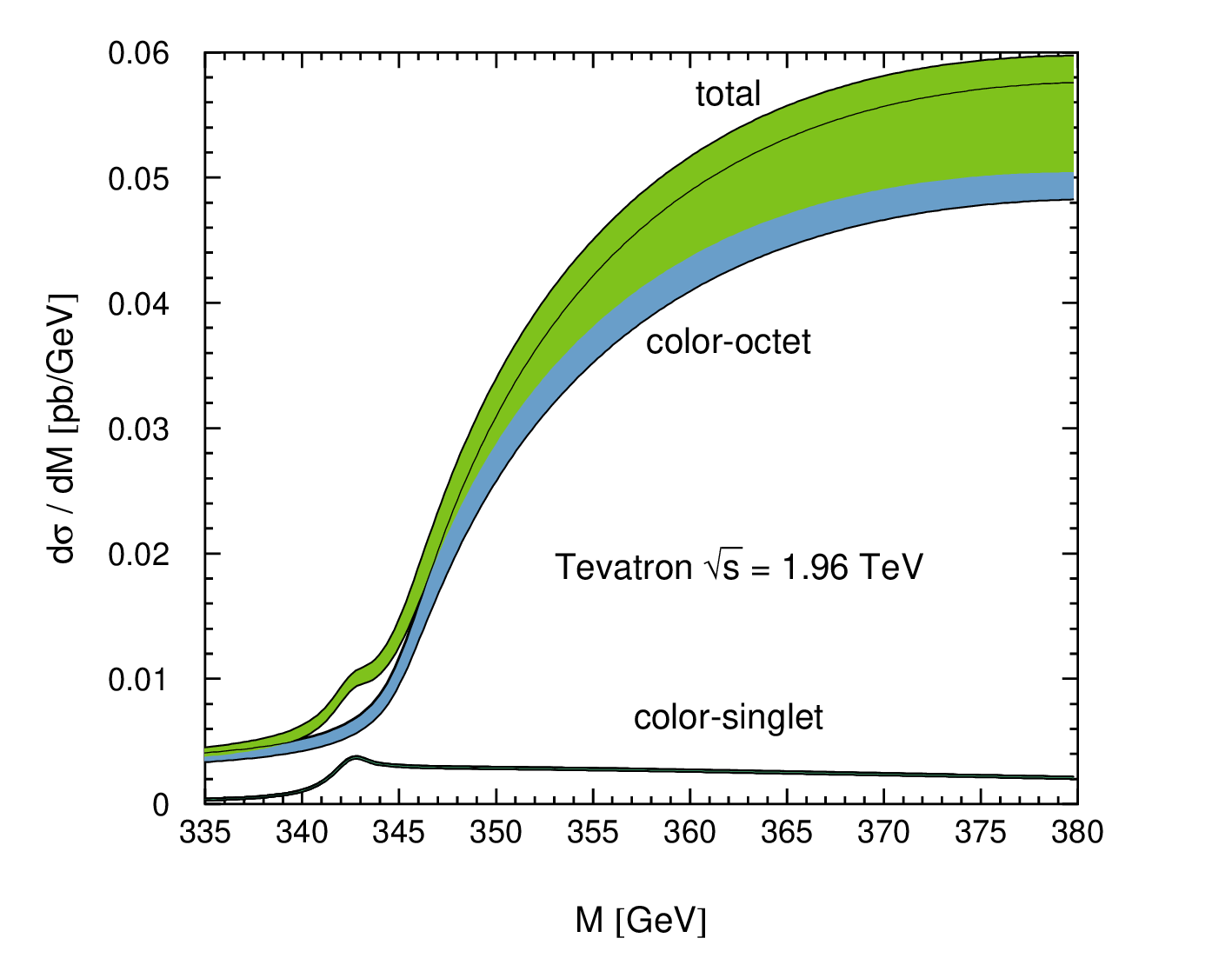}
\caption{
\label{fig:xs2tev}
Invariant mass distribution ${\rm d}\sigma/{\rm d}M$ for Tevatron with
$\sqrt{s}=1.96~{\rm TeV}$. At the Tevatron $q\bar{q}\rightarrow
{^1S_0^{[8]}}$ dominates the cross section, and luminosity
for $gg$ channels is small, thus the boundstate peak is 
buried by color-octet production.
}
\end{center}
\end{figure}

Although the most detailed top quark studies will be performed at the
LHC at an energy of $14~{\rm TeV}$, a sample of top quarks has been
collected at the Tevatron in proton anti-proton collisions at 
$1.96~{\rm TeV}$. Furthermore the first LHC data set will be taken
at $10~{\rm TeV}$. For this reason we give the results for these two
cases, in Figs.~\ref{fig:xs2lhc10} and \ref{fig:xs2tev}.
{The cross section in Fig.~\ref{fig:xs2lhc10} has the same 
characteristic shape as the one in Fig.~\ref{fig:xs2}, however, the
absolute size is considerably smaller.
As expected, the enhancement at threshold is significantly less pronounced
for Tevatron where the colour singlet contribution is very small.
}

{
Our analysis confirms the findings of Ref.~\cite{Hagiwara:2008df}, however,
the numerical results for the cross sections as presented in
Fig.~\ref{fig:xs2} are 
slightly higher than the corresponding corrections of
Ref.~\cite{Hagiwara:2008df} which is due to the combined effect of the
soft-gluon resummation, the inclusion of the NLO sub-processes and the 
different matching to full QCD.
}


\section{Summary}
\label{sec::summary}

A NLO analysis of top quark production near threshold
at hadron colliders has been performed.
The large width of the top quark in combination with the large
contribution from gluon fusion into a (loose bound) color singlet 
$t\bar t$ system leads to a sizable cross section for masses of the 
$t\bar t$ system significantly below the nominal threshold.
A precise measurement of the $M_{t\bar t}$ distribution in this region
which is dominated by the color singlet configuration could lead to a
top-quark mass determination which does not involve the systematic
uncertainties inherent in the determination of the mass of a single
(colour triplet) quark.
{Furthermore, also the shape of the differential distribution
${\rm d}\sigma/{\rm d}M$ is distorted and the mean $\langle M\rangle$ shifted
towards smaller values.
}

The effects of initial state radiation as well as boundstate corrections
are taken into account in consistent manner at NLO. 
{As compared to Ref.~\cite{Hagiwara:2008df} we include the 
complete $\hat{s}$ dependence in the matching condition and
also implement all NLO sub-processes. We 
observe a partial numerical cancellation between these two
effects leading to similar predictions as
Ref.~\cite{Hagiwara:2008df}.
}
Furthermore we
perform a soft-gluon resummation and thus include the dominant
logarithmically enhanced higher order terms. This last step
stabilizes the prediction. However, it enhances the cross 
section at most by 10\%.

The effects are more pronounced at the LHC with top production being
dominated by gluon fusion and less relevant in proton-antiproton
collisions with top quarks dominantly in color octet states.
Considering the threshold region (say up to $M_{t\bar t}=350~{\rm GeV}$) 
seperately, an integrated cross section of 
{15~pb} is obtained, which should be compared to 
{5~pb} as derived from the NLO predictions
using a stable top quark and neglegting the binding correction.
Within this relatively narrow region the enhancement amounts to roughly
a factor three and a significant shift of the threshold. Compared to
the total cross section for $t\bar t$ production of 
about {840~pb (obtained using fixed-order NLO accuracy for $\mu=m_t$, see,
  e.g., Ref.~\cite{Moch:2008qy})}, 
the increase is relatively small, about 
1\%. However, in view of
the anticipated experimental precision of better than
10\% these effects should not be ignored. 


\subsection*{Acknowledgments}

This work was supported by the DFG through SFB/TR~9, by the BMBF
through contract 05HT4VKAI3 
and by the Helmholtz Gemeinschaft under contract VH-NG-105.


\subsection*{Note added}
While this article was finished an analytic evaluation
of the total cross section at NLO accuracy appeared~\cite{Czakon:2008ii},
which has been used in Ref.~\cite{Hagiwara:2008df} to clarify the existence of 
a non-decoupling top quark effect overlooked in Ref.~\cite{Petrelli:1997ge}
(see footnote~3  on page~73 in Ref.~\cite{Hagiwara:2008df}).


{\footnotesize{


}}


\end{document}